\documentclass{svmult}

\usepackage{psfrag}
\usepackage{makeidx}         
\usepackage{graphicx}        
\usepackage{color}
\definecolor{red2}{rgb}{0.8,0.0,0.0}

\usepackage{multicol}        
\usepackage[bottom]{footmisc}

\makeindex             

\begin{document}





%
%
%
%





\newcommand{\up}{\uparrow}
\newcommand{\dn}{\downarrow}
\newcommand{\bra}[1]{\langle #1|}
\newcommand{\ket}[1]{|#1\rangle}
\newcommand{\braket}[2]{\langle #1|#2\rangle}
\newcommand{\aver}[1]{\langle #1 \rangle}
\newcommand{\vek}[3]{ \left[ \begin {array}{c} #1\\ #2\\ #3 \end {array} \right]}
\newcommand{\diag}{\mathop{\mathrm{diag}}}
\newcommand{\Tr}{\mathop{\mathrm{Tr}}}



\author{Matthias Braun, J\"urgen K\"onig, and Jan Martinek}

\title*{Manipulating Single Spins in Quantum Dots Coupled to Ferromagnetic Leads}
\titlerunning{Manipulating Single Spins in QDs Coupled to Ferromagnetic Leads}
\author{Matthias Braun\inst{1}, J\"urgen K\"onig\inst{1}, and Jan Martinek\inst{2,3,4}}
\institute{Institut f\"ur Theoretische Physik III, Ruhr-Universit\"at Bochum, 44780 Bochum, Germany
\and
Institute of Molecular Physics, Polish Academy of Science, 60-179 Pozna\'n, Poland
\and
Institut f\"ur Theoretische Festk\"orperphysik, Universit\"at Karlsruhe, 76128 Karlsruhe, Germany
\and
Institute for Materials Research, Tohoku University, Sendai 980-8577, Japan}

%
%
\maketitle
We discuss the possibility to generate, manipulate,
and probe single spins in single-level quantum dots
coupled to ferromagnetic leads. The spin-polarized
currents flowing between dot and leads lead to a
non-equilibrium spin accumulation, i.e., a finite
polarization of the dot spin. Both the magnitude and
the direction of the dot's spin polarization depends
on the magnetic properties of leads and their coupling
to the dot. They can be, furthermore, manipulated by
either an externally applied magnetic field or an
intrinsically present exchange field that arises due
to the tunnel coupling of the strongly-interacting
quantum-dot states to spin-polarized leads.
The exchange field can be tuned by both the gate and
bias voltage, which, therefore, provide convenient
handles to manipulate the quantum-dot spin. Since the
transmission through the quantum-dot spin valve
sensitively depends on the state of the quantum-dot
spin, all the dynamics of the latter is reflected in
the transport properties of the device.

\section{Introduction}
\label{chap:introduction} 

The study of single spins in quantum-dot spin valves
resides in the intersection of the two highly-interesting
and extensively-pursued research fields of spintronics
on the one hand and transport through nanostructures
on the other hand side. Quantum dots consist of a small
confined island with a low capacity such that a macroscopic
gate or bias voltage is needed to add a single electron,
leading to Coulomb-blockade phenomena
\cite{chargereview1,chargereview2,chargereview3}.
The notion that not only the charge but, simultaneously,
also the spin degree of freedom of the electrons can be
made use of, for example by using ferromagnetic leads,
defines the field of spintronics \cite{spinreview1,spinreview2}.
A quantum-dot spin valve, i.e., a quantum dot coupled
to ferromagnetic leads, exploits both the spin
polarization of the electrons and the sensitivity of
the charge to Coulomb interaction. Therefore electronic
transport is governed by the behavior of a {\em single spin}.
To discuss the possibility to generate, manipulate, and
probe single spins via electronic transport through
quantum-dot spin valves is the goal of this chapter.

The capacity $C$ of a metallic or semiconductor island
decreases when shrinking its size. For small quantum dots,
the energy scale to add a single electron on the island,
the charging energy $U=e^2/2C$, exceeds the energy scales
set by temperature $k_{\rm B}T$ or bias voltage $eV$, and
Coulomb-blockade phenomena arise, as first observed by
Fulton and Dolan \cite{Fulton}.
If, in addition, the island size becomes comparable to the
Fermi wavelength then the level spectrum on the island will
be discrete. For sufficiently large energy-level spacings,
only a single level may participate in transport.
Such as system can then be described by the Anderson-impurity
model, which is introduced below.

Famous examples of spintronics devices are the spin valves
based on either the giant magnetoresistance effect \cite{GMR}
in magnetic multilayers or the tunnel magnetoresistance
\cite{Julliere} in magnetic tunnel junctions.
These effects arise, when two ferromagnetic leads are in
contact via a conducting layer or a tunnel barrier,
respectively. The transport characteristics of the device
then depend on the relative orientations of the lead
magnetizations. If, in a magnetic tunnel junction,
the lead magnetizations enclose the angle $\phi$, the
conductance through the tunnel junction is proportional to
$\cos \phi$ \cite{Slonczewski,Angular}, i.e., it is maximal
for parallel and minimal for antiparallel alignment of
the leads' magnetizations. This angular dependence simply
reflects the overlap of the spinor part of the majority-spin
wave functions in the source and drain electrode, which is
given by the externally controlled leads' magnetizations.

This picture changes once spin accumulation can occur.
Let us consider transport through a
ferromagnet -- nonmagnet -- ferromagnet sandwich structure
with the thickness of the normal layer being smaller than
the spin diffusion length. A finite bias voltage applied
between the two ferromagnets with nonparallel magnetization
directions leads to a local imbalance of spin-up and
spin-down electrons in the nonmagnetic layer. This
non-equilibrium polarization of the electrons in the
nonmagnetic region, known as spin accumulation, mediates
the information of the relative orientation of the leads'
magnetization through the middle part, such that the
transmission through the device is reduced for increasing
angle $\phi$ between directions of the leads' magnetic
moments.

\begin{figure}[ht]
\centering
\includegraphics[width=0.7\columnwidth]{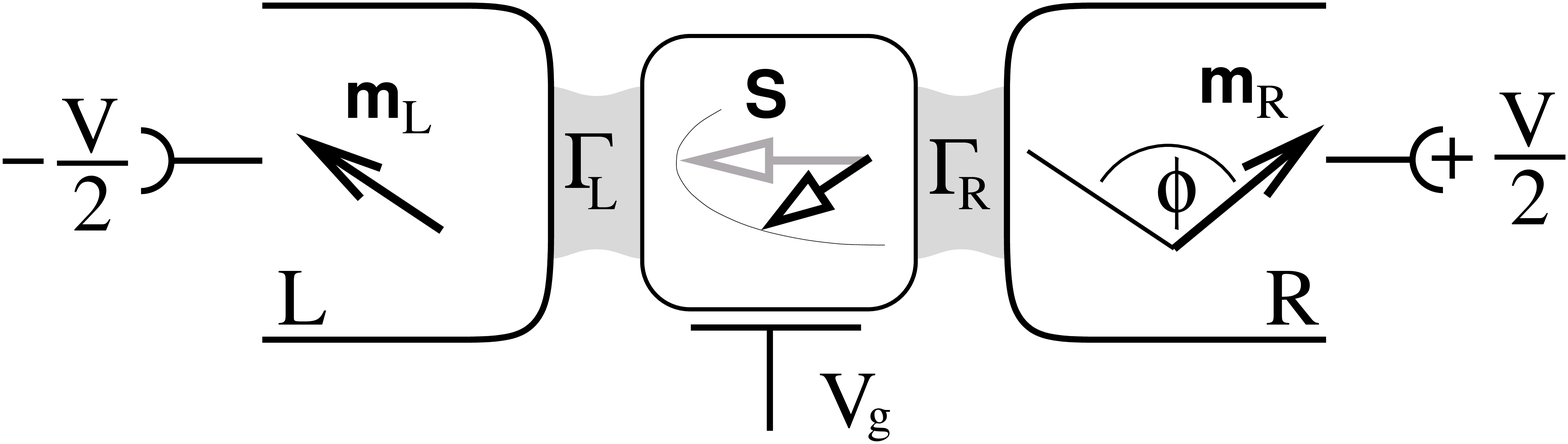}
\caption{\label{fig:system}A quantum dot contacted by
ferromagnetic leads with non-collinear magnetizations.
The lead magnetization directions enclose the angle
$\phi$. By forcing a current through the system, a
non-equilibrium spin $\vec{S}$ accumulates on the
otherwise non-magnetic dot.}
\end{figure}

An extreme limit of the above scheme is realized in a
quantum-dot spin valve. It consists of a quantum dot
that is tunnel coupled to ferromagnetic, see
Fig.~\ref{fig:system}.
In this case, the information about the relative leads'
magnetization directions is mediated by a single
quantum-dot electron that, as a consequence of a finite
bias voltage is partially spin polarized, described by
a finite quantum-statistical average
$\vec{S}= (\hbar/2) \aver{\vec{\sigma}}$ of the dot spin.
It is the orientation of the dot spin relative to the
leads together with the degree of the dot spin polarization
that determines the transport, rather than just the
relative orientation of the leads' magnetization
directions only. Any manipulation of the dot spin
polarization will change the transmission through the
device.

A quantum-dot spin valve is, thus, a convenient tool
to generate, manipulate, and detect spin polarization
of single quantum-dot electrons. Both the {\em generation}
and the {\em detection} of spin polarization on the
quantum dot occur via electrical transport as a
consequence of spin-polarized charge currents from and
to the leads. One of the intriguing features of a
quantum-dot spin-valve device is the possibility to
further {\em manipulate} the dot spin. This can be done
directly via an externally applied magnetic field \cite{EPL}.
But also the gate and transport voltages influence the
dot spin \cite{PRL,PRB,Braig,Usaj}. To
understand this, it is important to notice
that the strong Coulomb interaction on the quantum dot
yields many-body correlations. As for the spin degree of
freedom, the quantum-dot electrons are subject to
an exchange field that arises as a many-body effect due to
the tunnel coupling to spin-polarized leads.
This exchange field sensitively depends on the system
parameters such as the gate and bias voltage. The latter,
therefore, provide suitable handles to manipulate the
quantum-dot spin.

Combining ferromagnetic (typically metallic) leads with
quantum dots, which are usually semiconductor structures
is experimentally challenging. Recent experimental approaches
to such a quantum-dot spin valve involve metallic islands
\cite{ono,wees}, granular systems \cite{granular}, carbon
nanotubes \cite{cnt} as well as single molecules \cite{ralph2}
or self-assembled quantum dots \cite{QD-exp,ralph} coupled
to ferromagnetic leads. Another possible realization would
rely on contacting a surface impurity acting as quantum dot
with a spin-polarized STM tip \cite{FMSTM}.

Successful demonstration of tunnel magnetoresistance through
a strongly interacting system has been reported by
Sahoo {\em et al.} \cite{schoenenberger} in single-wall
carbon nanotubes contacted by PdNi leads, and by
Zhang {\em et al.} in Al grains sandwiched inside a
tunnel junction between to Co leads \cite{grain_experiment}.

The article is organized as follows:
In Sec.~\ref{sec:model} we define
the Hamiltonian of the quantum dot
coupled to ferromagnetic leads.
In Sec.~\ref{sec:dotdynamics} we address
the dynamics of the dot spin and charge.
Starting from a rigid calculation of the
spin and charge current through a tunnel
junction, we construct the master/Bloch
equation for the charge/spin degrees of freedom
from the charge/spin continuity equation.
From the Bloch equation, we discuss, how to
prepare, and modify the dot spin via bias
voltage, gate voltage and an external
applied magnetic field.
The so prepared dot spin can be measured by
it's imprint on the conductance of the quantum
dot spin valve device as shown in
Sec.~\ref{sec:manipulation}.
We summarize our findings then in
Sec.~\ref{sec:conclusions}.

\section{Model Hamiltonian}
\label{sec:model}
We describe the quantum-dot spin valve by
the following Hamiltonian \cite{PRL,PRB}:
\begin{eqnarray}\label{hamiltonian}
   H&=& \sum_{rk\alpha}
   \varepsilon^{\,}_{rk\alpha}c^{\dag}_{rk\alpha}c^{\,}_{rk\alpha}
   + \sum_{n}\varepsilon_n^{\,}
   d^{\dag}_{n}d^{\,}_{n} + U\,d^{\dag}_{\up}d^{\,}_{\up}
   d^{\dag}_{\dn}d^{\,}_{\dn}
\\ \nonumber &&
   +\sum_{rk\alpha n}
   \left(V_{rk\alpha n}c^{\dag}_{rk\alpha}d^{}_{n} + h.c.\right)\,.
\end{eqnarray}
The first term in Eq.~(\ref{hamiltonian}) treats the
ferromagnetic leads $(r=\rm L / R)$ as large reservoirs
of itinerant electrons. The Fermion creation and
annihilation operators of the lead $r$ are labeled by
$c^{(\dag)}_{rk\alpha}$, where $k$ labels the momentum
and $\alpha = \pm$ the spin. The spin-quantization axis
for the electrons in reservoir $r$ is chosen along its
magnetization direction $\vec{m}_{r}$. In the spirit of
the Stoner model, the property of ferromagnetism is
incorporated by assuming an asymmetry in the density of
states $\xi_\alpha$ for majority ($+$) and minority ($-$)
spins. 
The degree of spin polarization in lead $r$ is characterized by the ratio
$p_r=(\xi_{r+}-\xi_{r-})/(\xi_{r+}+\xi_{r-})$.
The lead
magnetization directions $\vec{m}_{\rm L}$ and
$\vec{m}_{\rm R}$ can enclose an arbitrary angle $\phi$.
Furthermore, the leads shall be so large, that the
electrons can always be described as in equilibrium, i.e.
with a Fermi distribution $f_r(\omega)$. An applied bias
voltage is taken into account by a symmetric shift of the
chemical potential in the left and right lead by $\pm eV/2$.

The quantum dot can be modeled as an Anderson impurity,
where $d^{\dag}_{n}$ and $d_{n}$ are the Fermion
creation and annihilation operators of the dot electrons
with the spin $n=\uparrow,\downarrow$.
The spin quantization axis of the dot is, in general,
chosen to be different from the quantization axes of both
the left and the right lead. If an external magnetic field
is applied, the spin quantization axis is chosen parallel to
this field. The energy $\varepsilon_n$ of the atomic-like
electronic level is measured relative to the equilibrium
Fermi energy of the leads, and double occupation
of the dot costs the charging energy $U\gg k_{\rm B}T$.

Electron tunneling between the leads and the dot is described
by the last term in Eq.~(\ref{hamiltonian}). As we have
chosen different spin quantization axes for the lead
subsystems, parallel to the respective magnetization, the
tunneling matrix elements $V_{rk\alpha n}$ are not diagonal
in spin space. However, we require that tunneling is spin
conserving. The tunneling amplitudes can, then, be separated
in $V_{rk\alpha n}=t_{rk}\,\times\,U^r_{\alpha n}$, i.e. a
spin-independent tunnel amplitude $t_{rk}$ and a $SU(2)$ rotation
matrix $U^r_{\alpha n}$. The explicit shape of the matrix is
determined by the choice of the dot spin quantization axis of
the dot system.

The tunnel-coupling strength is characterized by the transition
rates $\Gamma_{r\alpha}(\omega)= 2 \pi \sum_{k} |t^{}_{rk}|^2
\delta(\omega-\varepsilon_{rk \alpha})$. For simplicity, we
assume the density of states $\xi_{\alpha}$ and the tunneling
amplitudes $t_r$ to be independent of energy, which implies
constant tunneling rates $\Gamma_{r\alpha}$. The spin asymmetry
in the density of states in the leads yields spin-dependent
tunneling rates, which are related to the leads' spin polarization
by $p_r=(\Gamma_{r+}-\Gamma_{r-})/(\Gamma_{r+}+\Gamma_{r-})$.

Throughout this article we focus on the limit of dot-lead
coupling $\Gamma_{r\alpha} \ll k_{\rm B}T , eV$ when transport
is dominated by first-order tunneling, i.e., it is sufficient
to calculate all expressions for the charge and spin current up
to first order in $\Gamma$. This excludes the regimes of
second-order transport (cotunneling) in the Coulomb-blockade
region \cite{weymann} or the Kondo regime (see e.g.
Ref.~\cite{kondo}).

\section{Quantum-Dot-Spin Dynamics}
\label{sec:dotdynamics}

The tunnel coupling of the quantum-dot levels to
spin-polarized leads yields a transfer of angular
momentum across each of the tunnel junctions.
This, together with a change of angular momentum
due to an externally applied magnetic field, defines
the dynamics of the quantum-dot spin polarization.
The stationary value of the latter is determined by
balancing all currents of angular momentum.
As we discuss in detail below, the total {\em spin}
current, i.e., transfer of angular momentum from the
leads to the dot, consists of two qualitatively
different contributions. One is associated with the
fact that {\em charge} currents from or to a ferromagnet
is spin polarized. This current, thus, transfers angular
momentum {\em along} the magnetization directions of
ferromagnet or quantum-dot spin accumulation. There is,
however, also an additional contribution of transfer of
{\em perpendicular} angular momentum, which can be
expressed in terms of a many-body exchange field acting
on the quantum-dot electrons.

For a careful treatment of the total transfer of
angular momentum, we first present a rigid calculation
of the spin current through a single tunnel junction
\cite{MS} in terms of non-equilibrium Keldysh Green's
functions. This will help us to identify under which
circumstances spin currents with perpendicular components
will contribute. Afterwards, we specify our result to the
weak-coupling regime of a quantum-dot spin valve and derive
in this limit Bloch-like rate equations for the
quantum-dot spin.

The currents will be functions of the unknown
density matrix elements. To derive the stationary
density matrix in a non-equilibrium situation,
we set the change of charge and spin on the dot
equal zero. Then, the Bloch equation for the
spin and the continuity equation for the charge
degree of freedom from a system of master equations.

\subsection{Spin Current Expressed by means of Green's Functions }
\label{subsec:greensfunctions}

Our calculation of the spin current in this subsection
will be in close analogy to the derivation of the charge
current according to Meir and Wingreen \cite{MeirWingreen}.
Let us first consider the spin current through one, say
the left, tunnel barrier. For a clearer notation, we mostly
drop the lead index in this section. The spin current
$\vec{J}_{\rm L}=\aver{\hat{\vec{J}}_{\rm L}}$ from the lead
into the dot is defined by the negative of the time
derivative of the total lead spin $\hat{\vec{S}}_{\rm L}
=(\hbar/2)\sum_{k\alpha\beta} c^{\dag}_{k\alpha}
\vec{\sigma}^{}_{\alpha\beta} c^{\,}_{k\beta}$, where
$\vec{\sigma}^{}_{\alpha\beta}$ denotes the vector of
Pauli matrices. From the Heisenberg equation we get
\begin{eqnarray}\label{heisenberg}
\hat{\vec{J}}_{\rm L}= -\frac{d}{dt}\hat{\vec{S}}_{\rm L}=-\frac{1}{\I\hbar}\left[\hat{\vec{S}}_{\rm L},H\right]\,.
\end{eqnarray}
Making use of the Fermion commutation relation, we can
find the spin-current operator as
\begin{eqnarray}\label{operator}
\hat{\vec{J}}_{\rm L}^{}&=&-\frac{1}{2\I}
  \sum_{k\alpha\beta n}\,\,
  V_{k\alpha n}\vec{\sigma}^{\star}_{\alpha\beta}\,c^\dag_{k\beta}\,d_{n} - V_{k\alpha n}^\star \vec{\sigma}^{}_{\alpha\beta}\,d^{\dag}_{n}\,c^{}_{k\beta}\,.
\end{eqnarray}

By introducing the Keldysh Green's functions
$G^<_{n,k\beta}(t)=\I \aver{c^\dag_{k\beta}(0)\,d_{n}(t)}$,
we can write the expectation value of the spin current as
\begin{eqnarray}\label{current1}
\vec{J}_{\rm L}&=&\frac{1}{2}
  \sum_{k\alpha\beta n}\int\frac{d\omega}{2\pi}\left(
V_{k\alpha n}\vec{\sigma}^{\star}_{\alpha\beta}\,G^<_{n,k\beta}(\omega) - V_{k\alpha n}^\star\vec{\sigma}^{}_{\alpha\beta}\,G^<_{k\beta,n}(\omega)\right)\,.
\end{eqnarray}
Since the Green's functions obey the Dyson equations \linebreak
$ G^{<}_{k\alpha,n}=
\sum_{m} V_{k\alpha,m}^{}[\,g_{k\alpha}^{t}\,G_{m,n}^<
-g_{k\alpha}^{<}\,G_{m,n}^{\bar{t}}\,]$ and
$G^{<}_{n,k\alpha}=
\sum_{m}V_{k\alpha,m}^{\star}[\,g_{k\alpha}^{<}\,G_{n,m}^t
-g_{k\alpha}^{\bar{t}}\,G_{n,m}^{<}\,]$, we can replace
the Green's functions in Eq.~(\ref{current1}) with the dot
Green's functions $G^<_{n,m}(t)
=\I \aver{d^\dag_{m}\,d_{n}^{}(t)}$ and the
free Green's functions of the lead. The latter are given by
$g^{<}_{k\alpha}= 2\pi \I f^+_{\rm L}(\omega)\delta(\omega-\varepsilon_{k\alpha})$,
$g^{>}_{k\alpha}= -2\pi \I f^-_{\rm L}(\omega)\delta(\omega-\varepsilon_{k\alpha})$,
$g^{\rm ret}_{k\alpha}= 1/(\omega-\varepsilon_{k\alpha}+\I 0^+)$,
and $g^{\rm adv}_{k\alpha}=\left(g^{\rm ret}_{k\alpha}\right)^\star$.
Here, $f^+_{\rm L}$ stands for the Fermi distribution function
in the lead $\rm L$ and $f_{\rm L}^-=1-f_{\rm L}^+$.

If we choose the dot spin quantization axis parallel
to the lead magnetization we can substitute the
tunnel matrix elements by
$V_{k\alpha,n}=t_{k}\,\delta_{\alpha n}$.
After a lengthy but straightforward calculation,
the spin current can be written as
\begin{eqnarray}\label{eq:finalspincurrent1}
  \vec{J}^{}_{\rm L}
  &=&\!\frac{\I}{4}
  \sum_{m,n}\int \frac{d\omega}{2\pi}\,\,\,\,
      \vec{\sigma}^{}_{m n}(\Gamma_{m}+\Gamma_{n})\left[ f^+_{\rm L}(\omega)\,G^{>}_{n,m}+ f^-_{\rm L}(\omega)\,G_{n,m}^{<}\,\right]\nonumber\\
      &&+\vec{\sigma}^{}_{m n}(\Gamma_{m}-\Gamma_{n})\left[ f^+_{\rm L}(\omega)\,(G^{\rm ret}_{n,m}+G^{\rm adv}_{n,m})+\frac{1}{\I\pi}{\int}^\prime \!\!dE\,\frac{G_{n,m}^<(E)}{E-\omega}\,\right]\, ,
\end{eqnarray}
with the tunnel rates $\Gamma_{n}(\omega)= 2 \pi \sum_{k} |t^{}_{rk}|^2
\delta(\omega-\varepsilon_{rk \alpha}) \delta_{\alpha n}$.

This is the most general expression for the spin
current flowing through a tunnel barrier. Since
the Green's functions $G_{n,m}$ were not
specified during the calculation,
Eq.~(\ref{eq:finalspincurrent1}) holds also for
other electronic systems than single-level quantum
dots.

If the dot state is rotationally symmetric about
$\vec{m}_{\rm L}$, all dot Green's functions
$G^{}_{\sigma\sigma^\prime}$ non-diagonal in spin
space vanish. Only in this case, the
spin current is proportional to the difference
between charge current
$ I^{\up}_{\rm L}= \I (e/h) \int d\omega\,\,\Gamma_{\up}
[ f^+_{\rm L}(\omega)\,G^{>}_{\up\up}
+ f^-_{\rm L}(\omega)\,G_{\up\up}^{<}\, ] $
carried by spin-up electrons and charge
current $I^{\dn}_{\rm L}$ carried by
spin-down electrons,
\begin{eqnarray}\label{zcomponent}
\vec{J}_{\rm L}=J^{z}_{\rm L}\,\vec{e}_{z}&=&\frac{\hbar}{2 e}\left(I^{\up}_{\rm L}-I^{\dn}_{\rm L}\right)\,\vec{e}_{z}
\end{eqnarray}

If the dot system breaks this rotational
symmetry, for example due to spin accumulation
along an axis different from $\vec{m}_{\rm L}$,
the simple result of Eq.~(\ref{zcomponent}) is
no longer correct.
In such a situation, the second line in
Eq.~(\ref{eq:finalspincurrent1}) yields an
additional spin-current component, oriented
transversal to both, the magnetization of
the lead, and the polarization of the dot. This
spin-current component describes the exchange
coupling between lead and dot spin, causing both
to precess around each other. Since the lead
magnetization is pinned usually, only the dot spin
precesses like in a magnetic field.

Brataas {\em et al.} \cite{spinmix} showed,
that at normal metal -- ferromagnet interfaces,
incoming electrons, with a spin orientation
non-collinear to the magnetization direction,
may experience a rotation of the spin direction
during backscattering. This mechanism is described
by the so called spin-mixing conductance, and also
generates a transverse component of the spin current.

\subsection{Spin Current Between Ferromagnetic Lead and Quantum Dot}
\label{subsec:spincurrentleaddot}

We now specify the above expressions for a
quantum-dot spin valve for weak tunnel coupling.
Since the expression for the spin current
in Eq.~(\ref{eq:finalspincurrent1}) does already
explicitly depend linearly on the tunnel coupling
$\Gamma_{r\sigma}$, we only need the zeroth-order
Keldysh Green's functions of the dot system to
describe the weak-coupling regime. They are given by
\begin{eqnarray}
G^{>}_{\sigma\sigma} (\omega) &=&-2 \pi \I P_{\bar{\sigma}} \delta(\omega - \varepsilon - U) -2 \pi \I P_{0} \delta(\omega - \varepsilon)\\
G^{<}_{\sigma\sigma} (\omega) &=&2 \pi \I P_{\sigma} \delta(\omega - \varepsilon)
+2 \pi \I P_{d} \delta(\omega - \varepsilon - U)\\
G^{>}_{\sigma\bar{\sigma}} (\omega) &=&2 \pi \I P^{\sigma}_{\bar{\sigma}} \delta(\omega - \varepsilon -U)\\
G^{<}_{\sigma\bar{\sigma}} (\omega) &=& 2 \pi \I P^{\sigma}_{\bar{\sigma}} \delta(\omega - \varepsilon)\\
G^{\rm ret}_{\sigma\bar{\sigma}} (\omega) &=&\frac{P^\sigma_{\bar{\sigma}}}{\omega-\varepsilon+\I0^+}+
\frac{P^\sigma_{\bar{\sigma}}}{\omega-\varepsilon+U+\I0^+}
\,=\, \left(G^{\rm adv}_{\sigma\bar{\sigma}} (\omega)\right)^\star\,
\end{eqnarray}
where $P^\chi_\eta$ are the matrix elements of
the reduced density matrix of the dot system,
\begin{eqnarray}
  \label{dotdm}
  \rho_{\rm dot}= \left(
  \begin{array}{cccc}
    P_0 &     0 &     0 &     0 \\
    0     & P_{\uparrow} & P^{\uparrow}_{\downarrow} &     0 \\
    0     & P^{\downarrow}_{\uparrow} & P_{\downarrow} &     0 \\
    0     &     0 &     0 & P_{\rm d}
  \end{array} \right) \, .
\end{eqnarray}
The diagonal, real entries $P_\chi \equiv P_\chi^\chi$
are the probabilities to find the dot in the
state empty $(0)$, occupied with a spin up
$(\uparrow)$ or down $(\downarrow)$ electron,
or double occupied (d) with a spin singlet. The
zeros in Eq.~(\ref{dotdm}) in the off diagonals
are a consequence of the total-particle-number
conservation. The inner $2\times2$ matrix is the
$SU(2)$ representation of the dot spin. The reduced
density matrix contains five independent parameters.
For convenience, we describe the quantum dot state by
the probabilities for the three charge states
$P_0$, $P_1=P^{\uparrow}_{\uparrow}
+P^{\downarrow}_{\downarrow}$, and $P_{\rm d}$
(with the normalization condition $P_0+P_1+P_{\rm d}=1$),
and the average-spin vector $\vec{S}=(P^{\uparrow}_{\downarrow}+P^{\downarrow}_{\uparrow},
\I P^{\uparrow}_{\downarrow}-\I P^{\downarrow}_{\uparrow},
P^{\uparrow}_{\uparrow}-P^{\downarrow}_{\downarrow})/2$.

Similarly to deriving the expression for the spin current, we can get 
a formula for the charge current $I_r$ through tunnel contact $r$ as
\begin{eqnarray}
I_r&=&-\frac{e}{h}
  \sum_{k\alpha n}\int d\omega \left(
V_{rk\alpha n} G^<_{n,rk\alpha}(\omega) - V_{rk\alpha n}^\star 
G^<_{rk\alpha,n}(\omega)\right)\,.
\end{eqnarray}
After choosing a spin-quantization axis for the dot spin and making use of the
Dyson equation for the Green's functions, we can plug in the 
dot Green's function given above to obtain the result 
\begin{eqnarray}
\label{eq:chargecurrent}
I_r&=& \Gamma_{r}\frac{2(-e)}{\hbar} \Biggl[
f^+_{r}(\varepsilon) P_0 + \frac{f^+_{r}(\varepsilon+U)-f^-_{r}(\varepsilon)}{2}P_1
- f^-_{r}(\varepsilon+U) P_{\rm d}\nonumber \\
&&-p_r \left[ f^-_r(\varepsilon) + f^+_r(\varepsilon+U) \right]
    \vec{S}\cdot \vec{m}_r \,\Biggl] \,,
\label{currentS}
\end{eqnarray}
that, of course, is independent of the choice of the dot spin's quantization 
axis.
Here, we defined $\Gamma_r\equiv(
\Gamma_{r\uparrow}+\Gamma_{r\downarrow})/2$.

It is worth to mention, that the dot spin $\vec{S}$
influences the conductance via the scalar product
$(\vec{S}\cdot \vec{m}_r)$. Therefore the tunnel
magnetoresistance depends cosine like on the relative
angle enclosed by lead magnetization and spin
polarization, i.e. it just resembles the behavior of a
tunnel junction between two ferromagnetic leads
\cite{Julliere,Slonczewski,Angular}.

The first-order spin current, on the other hand, is given by
evaluating Eq.~(\ref{eq:finalspincurrent1}), which
leads to
\begin{eqnarray}
\label{eq:spincurrent}
\vec{J}_{r}&=&
\frac{\hbar}{2e}
I_r p_r \vec{m}_r
-\frac{\vec{S}-p_r^2(\vec{m}_r \cdot \vec{S})\vec{m}_r}{\tau_{{\rm c},r}}
       +\vec{S} \times \vec{B}_r\,.
\end{eqnarray}
The first term describes spin injection from the
ferromagnetic lead into the quantum dot by a
spin-polarized charge current.
The injected spin is proportional to the lead
polarization and the electrical current crossing
the junction. This spin current contribution
vanishes for vanishing bias voltage.

The second term describes relaxation of the dot
spin due to coupling to the leads. Since neither
an empty nor a doubly-occupied dot can bear a net
spin, the spin relaxation time $\tau_{{\rm c},r}^{-1}=
\Gamma_r/\hbar(1-f_r(\varepsilon) +f_r(\varepsilon+U))$
equals the life time of the single-occupation
dot state. This relaxation term is anisotropic
\cite{bauer}. The spin polarization of the lead
suppresses the relaxation of a dot spin, which is
aligned parallel to the lead magnetization.

The third term in Eq.~(\ref{eq:spincurrent})
describes transfer of angular momentum perpendicular
to the spin-polarization directions of lead and dot.
The structure of this terms suggests the
interpretation of $\vec{B}_r$ as being an effective
magnetic field that acts on the quantum-dot spin
$\vec{S}$. Its value, in the absence of an external
magnetic field, is given by \cite{PRL,Springer}
\begin{equation}
\label{eq:exchange}
  \vec{B}_r = p_r \,\frac{\Gamma_r{\bf \hat n}_r}{\pi\hbar} \int'
  d\omega \left( \frac{f^+_r(\omega)}{\omega-\varepsilon-U}
    +\frac{f^-_r(\omega)}{\omega-\varepsilon} \right) \, ,
\end{equation}
where the prime at the integral indicates Cauchy's
principal value. From Eq.~(\ref{eq:exchange}) it is
clear that this field is an exchange field that
arises due to the fact that the quantum dot levels
are tunnel coupled to a spin-polarized lead.
It is a many-body effect as all degrees of freedom
in the leads contribute to the integral, and Coulomb
interaction in the dot is important not to cancel
the first with the second term in the integrand.
The exchange field persists also for vanishing bias voltage.
A signature of this exchange field in the Kondo-resonance
splitting of transport through a single molecule has
been observed recently \cite{ralph2}, with reported
values of the field of up to 70 Tesla.

\subsection{Dynamics of the Quantum-Dot Spin}

We use the expressions for the charge and spin
current, Eqs.~(\ref{eq:chargecurrent}) and
(\ref{eq:spincurrent}), to calculate the dynamics
of the dot's charge and spin.
Strictly speaking, above calculation holds only
for static systems. To emphasize the physical origin
of the following equations, we keep all time
derivatives in this section, even if they should
have the numerical value of zero.

The continuity equation of the average dot charge
$\aver{n}=\sum_n n P_n$ is given by
\begin{eqnarray}
\label{eq:chargemaster}
e \frac{d \aver{n}}{dt}&=& I_{\rm L} + I_{\rm R}\,.
\end{eqnarray}
Moreover, not only the total charge current
through the dot is conserved, but also the charge
current through the individual charge levels.
Therefore we can split the charge continuity
Eq.~\ref{eq:chargemaster}
into the two contributions associated with transport
processes in which either double occupied or an empty
dot is involved.
The affiliation to either contribution is indicated by
the arguments of the Fermi functions, where the presence
of the interaction energy $U$ indicates processes with
double occupation and the absence signals processes
involving an empty dot. We get
\begin{eqnarray}\label{eq:P0d}
\frac{d P_0}{dt}&=&\sum_{r}\Gamma_{r} \Biggl( \,
f^+_{r}(\varepsilon) P_0
-f^-_{r}(\varepsilon)P_1/2
-p_r f^-_r(\varepsilon) \vec{S}\cdot \vec{m}_r \,\Biggl)\qquad\\
\frac{d P_{\rm d}}{dt}&=& \sum_{r}
\Gamma_{r} \Biggl( \,f^+_{r}(\varepsilon+U)P_1/2
- f^-_{r}(\varepsilon+U) P_{\rm d}-p_r f^+_r(\varepsilon+U) \vec{S}\cdot \vec{m}_r\, \Biggl)\, .\nonumber
\end{eqnarray}

Similar to the charge continuity equation,
the continuity equation for the dot spin reads
\begin{eqnarray}
\label{eq:spinmaster}
\frac{d\vec{S}}{dt}&=& \vec{J}_{\rm L} + \vec{J}_{\rm R}
+ \vec{S} \times \vec{B}_{\rm ext} - \frac{\vec{S}}{\tau_{\rm rel}}
\nonumber \\
&=&\frac{\hbar}{2e} \sum_{r} \left[
I_r p_r \vec{m}_r
-\frac{\vec{S}-p^2(\vec{m}_r \cdot \vec{S})\vec{m}_r}{\tau_{{\rm c},r}} \right]
     +\vec{S} \times \vec{B}_{\rm tot} - \frac{\vec{S}}{\tau_{\rm rel}} \,.
\end{eqnarray}
with $\vec{B}_{\rm tot} = \left( \vec{B}_{\rm L} + \vec{B}_{\rm R} +
\vec{B}_{\rm ext} \right)$.
In addition to the spin currents entering the
quantum dot from the left and right lead, there
is one term describing spin precession due to an
external magnetic field $\vec{B}_{\rm ext}$.
It enters the equation in the same way as the
exchange field with the left and right reservoir,
so that all three of them add up to the total field
$\vec{B}_{\rm tot}$.
Furthermore, we phenomenologically took into
account the possibility of intrinsic spin
relaxation, e.g., due to spin-orbit
coupling, hyperfine interaction with nuclei in
the quantum dot, or higher-order tunnel processes
such as spin-flip cotunneling, with a time scale
$\tau_{\rm rel}$. The total spin-decoherence time
of the dot spin is, therefore, given by
\begin{eqnarray}\label{tau}
  \left(\tau_{\rm{s}}\right)^{-1} = \left(\tau_{\rm{rel}}\right)^{-1}
  + \left(\tau_{\rm{c, L}}\right)^{-1} + \left(\tau_{\rm{c, R}}\right)^{-1}\, .
\end{eqnarray}

\begin{figure}[ht]
\centering
\includegraphics[width=0.7\columnwidth]{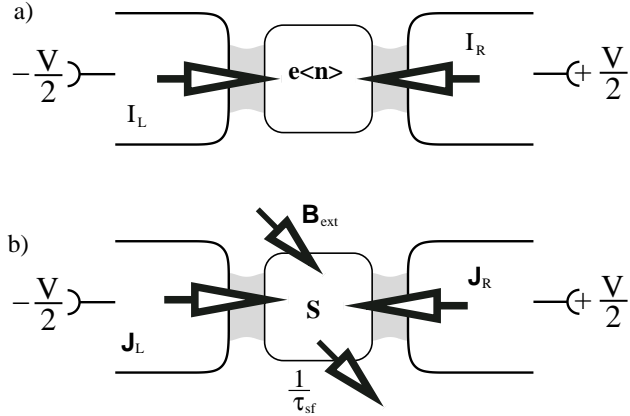}
\caption{\label{fig:master}
    a) The dot charge changes according to the
    electrical current through the tunnel barriers.
    b) The dot spin changes according to the spin
    currents through the tunnel barriers. In addition,
    an external magnetic field acts as additional
    source and the intrinsic spin
    relaxation as sink of angular momentum.}
\end{figure}

The different handles to manipulate the
quantum-dot spin are comprised the
total field $\vec{B}_{\rm tot}$.
It contains the external magnetic field
$\vec{B}_{\rm ext}$ as a direct tool
to initiate a spin precession.
However, also the the exchange fields can
be used for this task \cite{PRB}. As we see from
Eq.~(\ref{eq:exchange}), the exchange field
depends on both the gate and bias voltage
via the level position $\varepsilon$ and the
Fermi distribution functions $f_r(\omega)$.

\section{Manipulation and Detection of the Quantum-Dot Spin via Electrical Transport}
\label{sec:manipulation}

Since the spin state of the quantum dot enters
the expressions for the charge current in
Eq.~(\ref{eq:chargecurrent}), any manipulation
on the quantum-dot spin can be detected in
measuring the $dc-$charge current through the device.

In order to calculate the charge current, we need
to determine the stationary solution for the density
matrix, i.e. for the dot spin $\vec{S}$
and the charge occupation probabilities $P_i$.
For these six variables, we need six
independent equations: the probability
normalization condition $\sum_n P_n=1$, the Bloch
equation $d\vec{S}/dt=0$ (contains three equations),
and the two equations originating from the
charge continuity.

In the following we analyze stationary transport
situations, i.e., neither the charge nor the spin
of the dot changes with time, and the currents
through the left and right tunnel junction are equal
$I_{\rm L}=-I_{\rm R}\equiv I$.
For simplicity, we choose symmetric coupling
$\Gamma_{\rm L} = \Gamma_{\rm R} = \Gamma/2$, equal
spin polarizations $p_{\rm L} = p_{\rm R} = p$, and
a symmetrically applied bias
$V_{\rm R} = -V_{\rm L} =  V/2$ for the following
discussion.

In the following three subsection we consider
the effect of the gate and transport voltage
as well as an external magnetic field on the
quantum-dot spin, and how this is reflected by
electric transport through the quantum-dot
spin valve.
Then, in reversal, by experimentally
measuring the transport characteristics of
the device, one can conclude the spin state
of the quantum dot.

\subsection{Gate Voltage Effect in Linear Response Regime}
\label{subsec:linearresponse}

\begin{figure}[!ht]
\begin{center}
\includegraphics[width=0.6\columnwidth,angle=0]{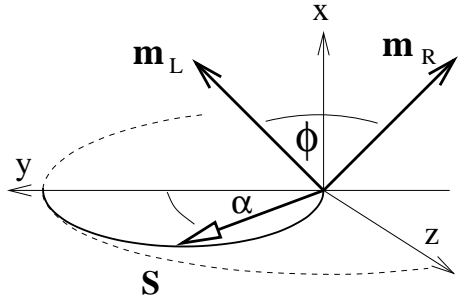}
\caption{
\label{fig:lineardynamics}
   Spin dynamics in the linear-response regime. Spin accumulates along
   the $y$ direction. The spin precesses due to the exchange field that is
   along the $x$ direction.
   Therefore, the stationary solution of the average spin on the dot is tilted
   away from the $y$ axis by an angle $\alpha$, plotted
   in Fig.~\ref{fig:linrescurrentplots}(b).}
\end{center}
\end{figure}

To study the effect of the gate voltage on the
quantum-dot spin via the gate-voltage dependence
of the exchange field, we analyze the linear-response
regime in the absence of an external magnetic field,
and for simplicity without intrinsic spin relaxation.
Without any applied bias voltage $V=0$, i.e. in
equilibrium the stationary solution of the rate
equations for the charge occupation probabilities
(\ref{eq:P0d}) is given by the Boltzmann
distribution, $P_\chi \sim \exp(- E_\chi/k_{\rm B}T)$,
and no current flows. Since the dot itself is non-magnetic,
the dot spin vanishes $\vec{S} =0$.
For a small bias voltage $eV \ll  k_{\rm B} T$, we can expand
the master Eq.~(\ref{eq:spinmaster}) and (\ref{eq:P0d})
up to linear order in $V$.
With symmetric couplings to the left and right lead,
the charge probabilities $(P_0,P_1,P_{\rm d})$ become
independent on $V$, thus, the occupation probabilities
are given by their equilibrium value, but the spin
degree of freedom is not. The linear charge current,
which is polarized due to the lead magnetizations
generate a finite dot-spin polarization along
$p(\vec{m}_{\rm L} - \vec{m}_{\rm R})$,
i.e., along the $y$ axis in the coordinate system
defined in Fig.~\ref{fig:lineardynamics}.

The damping term in Eq.~(\ref{eq:spinmaster}) limits the magnitude of
spin accumulation. The term $\vec{S} \times \left(
\vec{B}_{\rm L} + \vec{B}_{\rm R}\right)$
yields an intrinsic precession of the spin around the exchange field
$\vec{B}_{\rm L} + \vec{B}_{\rm R}\equiv  B_0 \cos(\phi/2)\, \vec{\rm e}_x$.
In the steady state, the average dot spin is rotated by the angle
\begin{equation}
  \alpha = - \arctan \left( B_0 \tau_{\rm s} \cos \frac{\phi}{2} \right)\,.
\end{equation}
Therefore the accumulated spin acquires both $y$ and $z$ components as
seen in Fig.~\ref{fig:lineardynamics}.
This precession also reduces the magnitude of the accumulated spin to
\begin{equation}
|\vec{S}| =  p\,I\, \tau_{\rm s}\,  \cos \alpha\,.
\end{equation}

\begin{figure}[!ht]
\begin{center}
\includegraphics[width=0.7\columnwidth,angle=0]{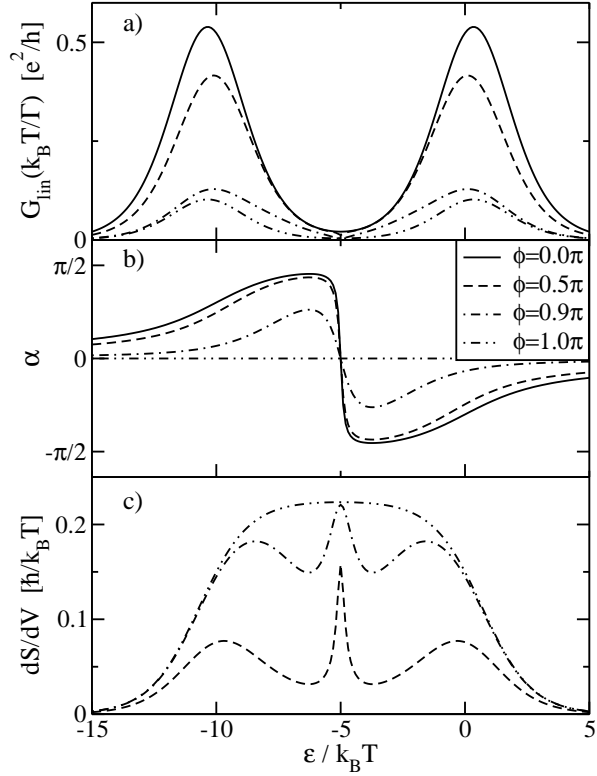}
\caption{\label{fig:linrescurrentplots} (a) Linear
conductance normalized by $\Gamma/k_{\rm B}T$ as a function of the level
position $\varepsilon$ for different angles $\phi$. (b) Angle $\alpha$ enclosed by the
accumulated spin and the $y$ axis as defined in
Fig.~\ref{fig:lineardynamics}. (c) Derivation of the magnitude of the accumulated spin on the dot with respect to the
source-drain voltage $V$. Further parameters are $p=0.9$ and $U=10k_{\rm B}T$.}
\end{center}
\end{figure}

The precession angle $\alpha$ is plotted in Fig.~\ref{fig:linrescurrentplots}(b)
as function of the level position $\varepsilon$, that can be tuned by the
gate voltage.
The angle $\alpha$ changes its sign at $\varepsilon = -U/2$, due to a
sign change of the exchange field at this point.
The level position $\varepsilon = -U/2$ is special, since then
the particle and hole like processes generating the exchange
field compensate each other.

\begin{figure}[!ht]
\begin{center}
\includegraphics[angle=-90,width=0.7\columnwidth]{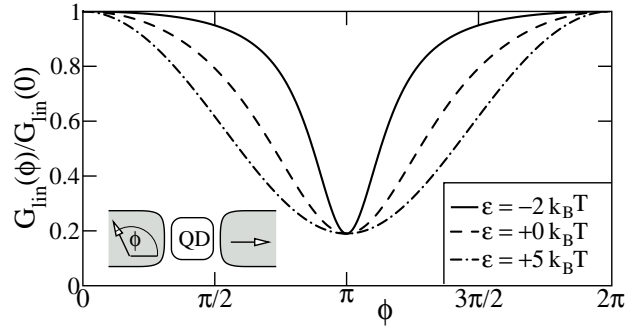}
\caption{\label{fig:normalizedconductance} Normalized conductance as
a function of the angle $\phi$ enclosed by the lead magnetization
for different level positions and the parameters $U=10k_{\rm B}T$ and $p=0.9$.}
\end{center}
\end{figure}

As pointed out above, in the linear-response regime under consideration
the charge occupation probabilities do not depend on the spin polarization
of the leads. In particular, they are independent of the relative angle
$\phi$ of the leads' magnetization.
This means that the $\phi$ dependence of the conductance is
determined by the product
$\vec{S}\cdot\vec{m}_{\rm L} = - \vec{S} \cdot \vec{m}_{\rm R}$,
as can be seen from Eqs.~(\ref{eq:chargecurrent}).
It is the relative orientation of the accumulated spin and the drain (or
source) that produces the $\phi$ dependence of the current, rather than the
product $\vec{m}_{\rm L} \cdot \vec{m}_{\rm R}$, as in the
case of a single magnetic tunnel junction.
Therefore the $\phi$ dependent linear conductance $G^{\rm lin} =
(\partial I/ \partial V)|_{V=0}$ directly reflects the accumulated spin.
The effect of the exchange field for the normalized conductance is seen
from the analytic expression
\begin{eqnarray}
\label{currlin}
   \frac{G^{\rm lin}(\phi)}{G^{\rm lin}(0)} &=& 1
   - p^2\frac{\sin^2 (\phi/2)}{1+ (B_0 \tau_{\rm s})^2 \cos^2(\phi/2)}
   \, , \qquad
\end{eqnarray}
which is plotted in Fig.~\ref{fig:normalizedconductance} for different
values of the level position $\varepsilon$.

For $\varepsilon > 0$, the quantum dot is predominantly empty, and
for $\varepsilon+U < 0$ doubly occupied with a spin singlet.
In this regions, the life time of a singly-occupied dot $\tau_{\rm c}$
is short, and so is the lifetime of the dot spin. Therefore the
rotation angle $\alpha$ is small and the normalized conductance as a function
of the relative angle $\phi$ of the lead magnetizations shows a
harmonic behavior, see, e.g., the curve for $\varepsilon = 5 k_{\rm B}T$ in
Fig.~\ref{fig:normalizedconductance}.

For $-U  < \varepsilon < 0$ the dot is primarily singly occupied, 
so the spin dwell time is increased and the exchange field becomes important.
It causes the above described spin precession, which decreases the product
$\vec{S} \cdot \vec{m}_{\rm L}$ since the relative angle
between $\vec{m}_{\rm L}$ and $\vec{S}$ is increased and the magnitude
of $\vec{S}$ is reduced.
Thus, the spin precession makes the spin-valve effect less pronounced,
leading to a value of the conductance that exceeds the expectations
made by Slonczewski in Re.~\cite{Slonczewski} for a single magnetic
tunnel junction.

For parallel and antiparallel aligned lead magnetizations, $\phi=0$ and
$\phi=\pi$, the accumulated spin and the exchange field also get aligned.
In this case, the spin precession stops, even though the exchange field
is still present. The $\phi$-dependent conductance is not affected by the
exchange field at this alignment, see Fig.~\ref{fig:normalizedconductance}.

\subsection{Bias Voltage Effect in Non-Linear Regime}
\label{subsec:continuity}

We now turn to the non-linear response regime, $eV>k_{\rm B}T$, in order to
discuss the effect of the bias-voltage dependence of the exchange field
on the quantum-dot spin.
Again, we assume that there is no external magnetic field, and no
spin relaxation.

\begin{figure}[!ht]
\begin{center}
\includegraphics[width=0.7\columnwidth,angle=0]{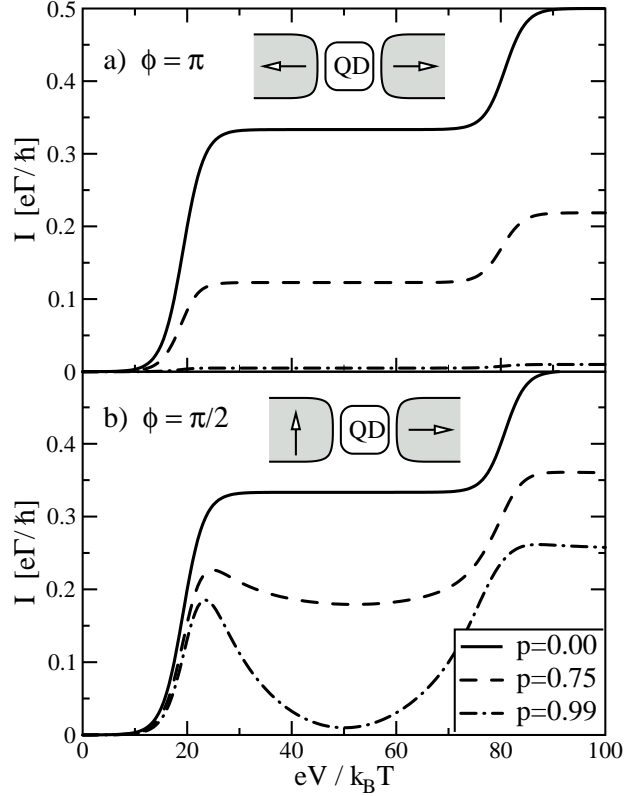}
\caption{\label{fig:currentplots} Current-voltage characteristics for
antiparallel (a) and perpendicular aligned (b) lead
magnetizations. Further
parameters are $\Gamma_{\rm L}=\Gamma_{\rm R}=\Gamma/2$, $p_{\rm L}=p_{\rm R}=p$, $\varepsilon=10{k_{\rm B}T}$, and
$U=30{k_{\rm B}T}.$}
\end{center}
\end{figure}

In Fig.~\ref{fig:currentplots}(a) we show the current $I$ as a function of the bias
voltage $V$ for an antiparallel configuration of the leads' magnetizations
and different values of the leads' spin polarization $p$.

For non-magnetic leads, the current-voltage characteristic
shows the usual Coulomb staircase.
At low bias voltage, the dot is empty and transport is blocked.
With increasing bias voltage, first single and then double occupancy of the
dot is possible, which opens first one and then two transport channels.
A finite spin polarization $p$ leads to spin accumulation and, thus, to a
reduction of transport.
A reduction of transport with increasing $p$ is also seen for noncollinear
magnetization. But there is a qualitative difference as can be seen in
Fig.~\ref{fig:currentplots}(b). A very pronounced negative
differential conductance evolves out of the middle plateau as
$p$ is increased. To understand the negative differential
conductance we first neglect the exchange field and then, 
in a second step, analyze how the exchange field modifies the picture.

At the intermediate bias voltages, the dot can only be empty or singly occupied,
double occupation is forbidden. Therefore all electrons entering
the dot through the left barrier find an empty dot. In this regime the current
$I = (e\Gamma/\hbar) P_0$ explicitly depends only on the probability to find
the dot empty. The transport through the dot must be suppressed by
charge accumulation on the dot, i,e, $P_1\rightarrow 1$ and $P_0\rightarrow 0$. 

The origin of this charge accumulation becomes clear from the relation of charge and spin
\begin{eqnarray}\label{eq:spinnonlinear}
\vec{S} \!\!&=&\!\! p \left[ \frac{\Gamma_{\rm L}}{\Gamma_{\rm R}}P_0 \vec{m}_{\rm L} - \frac{1-P_0}{2}\,\vec{m}_{\rm R} \right]\,.
\end{eqnarray}
If, in the steady state, the dot is primarily occupied by one electron, 
this electron has a spin, which is antiparallel aligned to the drain lead. 
Due to this antiparallel alignment, the tunneling rate to the drain lead is
maximally suppressed, while the tunnel coupling to the
source lead is not as much affected. When the rate to the drain lead is weak,
but strong to the source lead, then the dot is primarily occupied by one
electron.

\begin{figure}[!ht]
\begin{center}
\includegraphics[width=0.6\columnwidth,angle=0]{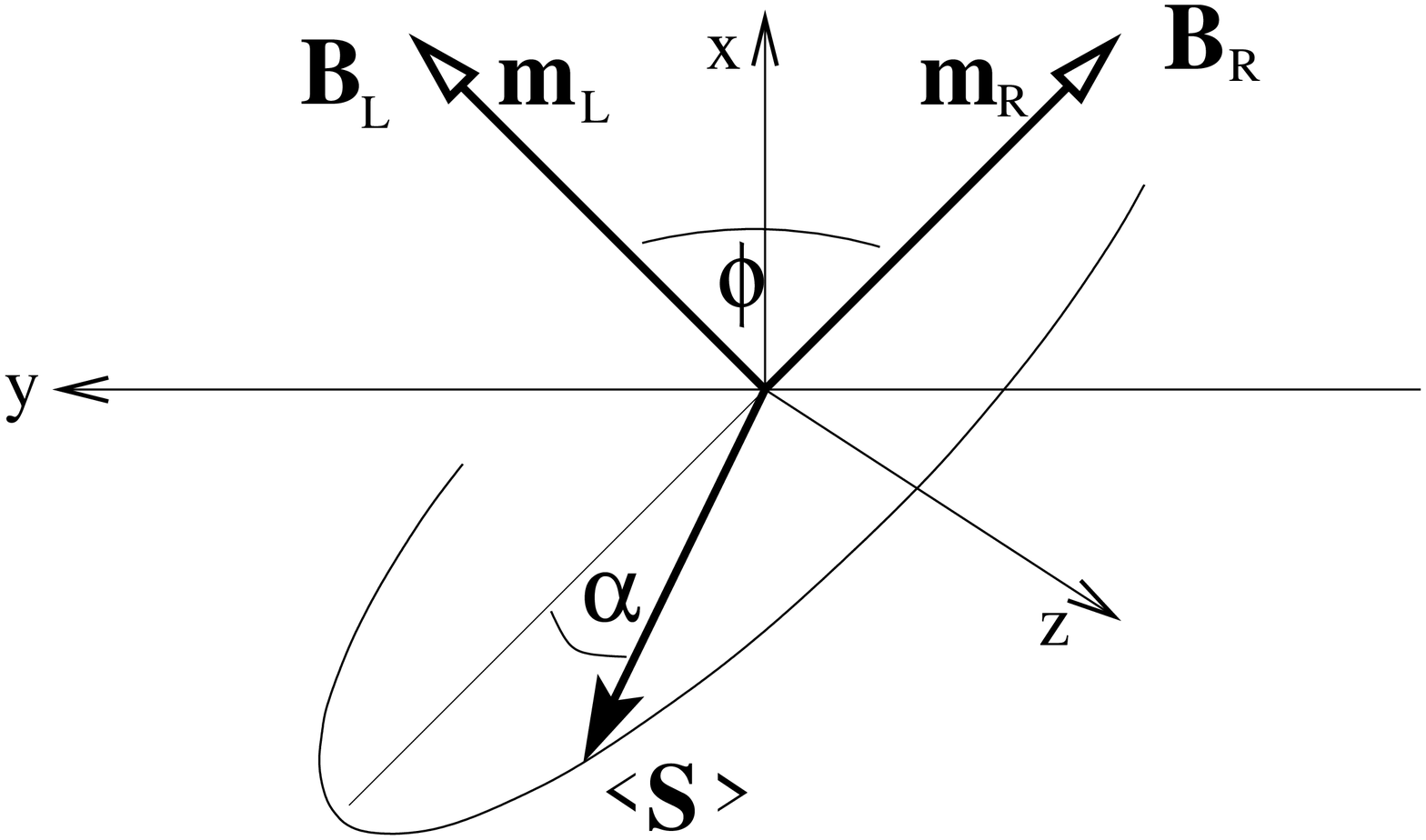}
\caption{\label{fig:dinlr} For electrons polarized antiparallel to
the drain lead, the influence of the effective field generated by
the source lead is dominating. By rotating the spins, the spin
blockade is lifted and therefore the conductance recovers.}
\end{center}
\end{figure}

The transport is suppressed, since an electron is trapped due to its spin alignment,
and no second electron can enter the dot because of the Coulomb interaction.
So this mechanism is a type of spin blockade but with a different physical
origin compared to the systems described in literature \cite{spinblock1,spinblock2}.
The suppression defines the local minimum of the current in Fig.~\ref{fig:currentplots}(b).
At this point, the relevant exchange field component generated by the coupling
to the left lead vanishes, so that spin precession becomes insignificant. 
Away from this point spin precession sets in as  illustrated in Fig.~\ref{fig:dinlr}.
The spin rotates about $\vec{m}_{\rm L}$,  the spin blockade gets lifted,
and the electron can now more easily leave the dot via the drain electrode.

\begin{figure}[!ht]
\begin{center}
\includegraphics[width=0.7\columnwidth,angle=0]{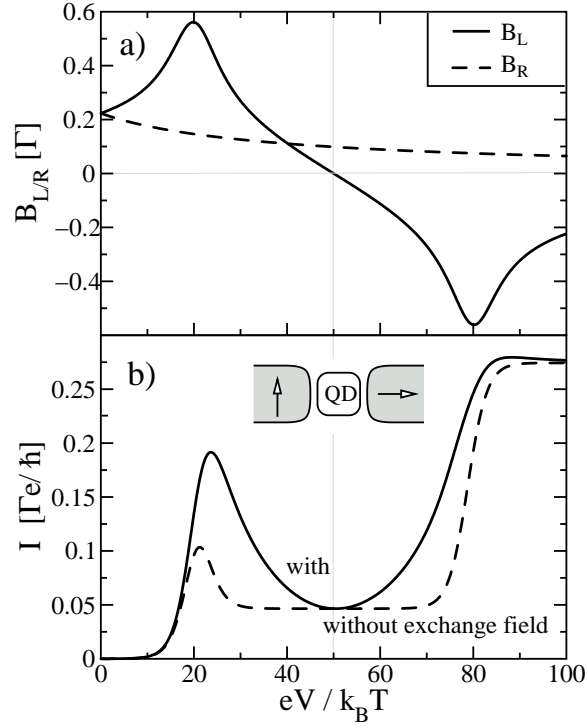}
\caption{\label{fig:exfield} Panel (a) The absolute value of the
effective exchange field contributions from the left and right
leads. Panel (b) the current voltage dependence, with and without
the influence of the exchange field. For both plots the parameters
$\phi=\pi/2$, $\Gamma_L=\Gamma_R=\Gamma/2$,
$\varepsilon=10k_{\rm B}T$, $U=30k_{\rm B}T$, and $p=0.95$ were chosen.}
\end{center}
\end{figure}

The particular value of the non-linear conductance is a consequence 
of the two competing effects. Spin blockade reduces, while spin precession again
increases the conductance. Since the strength of the exchange field varies
as a function of the level position with respect to the Fermi level, 
see Fig.~\ref{fig:exfield}(a), this recovery is non-monotonous,
what leads to a negative differential conductance.
To illustrate this further, we plot in Fig.~\ref{fig:exfield}(b) the current which
we obtained when the spin precession contribution is in an artificial way dropped in Eq.~(\ref{eq:spinmaster}), and compare it with the total current.
In the absence of the exchange field, a wide plateau is recovered, whose
height is similar to the current, one would expect, if the lead
magnetizations were aligned antiparallel.
The peak at the left end of the plateau indicates that, once the dot level
is close to the Fermi level of the source electrode, the spin blockade is
relaxed since the dot electrons have the possibility to leave to the
left side.

However, this negative differential conductance occur only at relatively
high values of the lead polarization. For symmetric tunnel coupling a
spin polarization of $p \approx 0.77$ is needed, while for a strong asymmetry
in the tunnel coupling the required spin polarization is reduced.

\begin{figure}[!ht]
\begin{center}
\includegraphics[angle=-90,width=0.7\columnwidth]{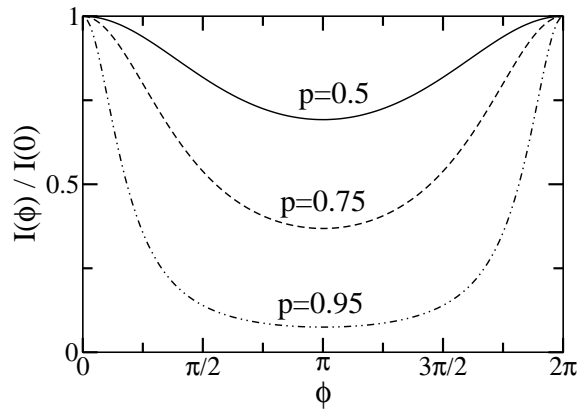}
\caption{\label{cond} Angular dependence of the conductance with
an applied voltage of $V=\varepsilon+U/2$, i.e., the voltage
generating the smallest influence of the exchange field. Further
plot parameters are $\Gamma_{\rm L}=\Gamma_{\rm R}=\Gamma/2$, $p_{\rm L}=p_{\rm R}=p$,
$\varepsilon=10k_{\rm B}T$, and $U=30k_{\rm B}T$.}
\end{center}
\end{figure}

The effect of the spin blockade on the $\phi$-dependent of the current is depicted
in Fig.~\ref{cond}.
We choose the bias voltage according to $eV/2=\varepsilon+U/2$, such that
the influence of the exchange field is absent.
For $p=0.5$ still a $\sin^2\frac{\phi}{2}$ dependence can be
recognized. For higher values of the spin polarization the
conductance drops faster and stays nearly constant at its minimal
value due to spin blockade.
This is just the opposite behavior than predicted for the
linear-response regime as seen in Fig.~\ref{fig:normalizedconductance}.

If such a high bias voltage is applied, that the dot
can also be double occupied, the step-like behavior
of the current voltage characteristic is recovered,
see Fig.~\ref{fig:currentplots}(b).
Away from the step, all appearing Fermi functions can
be approximated by $0$ or $1$, and following
Eq.~(\ref{eq:chargecurrent}) the current is given by
$I = (e\Gamma/2\hbar) \left[
1-p \vec{S} \cdot (\vec{m}_{\rm L} - \vec{m}_{\rm R}) \right]$.
Far away form the resonance, where the exchange field can
be neglected, the accumulated spin is
$\vec{S} = p (\vec{m}_{\rm L} - \vec{m}_{\rm R})/4$
from which we get
\begin{equation}
  I = \frac{e\Gamma}{2\hbar} \left( 1-p^2 \sin^2 \frac{\phi}{2} \right) \, .
\end{equation}
The suppression of transport due to the spin polarization $p$ of the leads is
comparable with the case of a single-tunnel junctions, when charging
effects are of no importance.

We close this section with the remark that while we plotted only results
for the case $\varepsilon >0$, in the opposite case $\varepsilon < 0$ the
current-voltage characteristics is qualitatively the same.

\subsection{External Magnetic Field}
\label{subsec:magnetic}

In the previous subsections were studied quantum-dot spin dynamics that is
evoked by the exchange field.
But one also make use of an externally-applied magnetic field
$\vec{B}_{\rm ext}$.
It turns out that with an external field one can measure the spin-decoherence
time $T_2$.
To emphasize this point, we explicitly allow intrinsic spin relaxation on the dot. Then, we observe a separation of the charge life time
$\tau_{\rm{c}} ^{-1} = \tau_{\rm{c, L}}^{-1} + \tau_{\rm{c, R}}^{-1}$ and the spin life time on the dot
$\tau_{\rm{s}}^{-1} = \tau_{\rm{c}} ^{-1}+\tau_{\rm{rel}}^{-1}$

An external field leads to the Hanle effect \cite{EPL}, {\it i.e.},
the decrease of spin accumulation in the quantum dot due
to precession about a static magnetic field.
Indeed, this was the effect used by Johnson and Silsbee \cite{johnson}
and others \cite{wees} to prove non-equilibrium spin accumulation.

Optical realizations of such Hanle experiments \cite{hanle_in_dots}
always involve an ensemble averaging over different dot realizations,
so the outcome of the measurement is $T_2^\star$ rather than $T_2$.
By measuring the Hanle signal via the conductance through a quantum
dot attached to ferromagnetic leads, this ensemble averaging is avoided.

In a recent experiment Zhang {\em et. al.} \cite{grain_experiment} realized
this kind of setup but with a whole layer of aluminum dots in a tunnel
junction between two Co electrodes. Even so the measurements involve averaging over
different realizations of the dots, multi levels and local magnetizations, they
clearly observe a Hanle resonance in the magnetoresistance of the device.

For simplicity we assume symmetric couplings $\Gamma_{\rm L}=\Gamma_{\rm R}$,
equal degree of lead polarizations $p_{\rm L}=p_{\rm R}=p$ and consider
the linear-response regime only.
There is a variety of possible relative orientations of the external field
and the leads' magnetizations to each other.
In the following, we consider two specific cases in detail, as they are
convenient to extract useful information about the spin-decoherence time
in one case, and to prove the existence of the exchange field in the other one.

\subsubsection{Antiparallel Aligned Lead' Magnetizations}
\label{subsubsec:magnetic}

\begin{figure}[h!]
\begin{center}
\includegraphics[angle=-90,width=1.0\columnwidth]{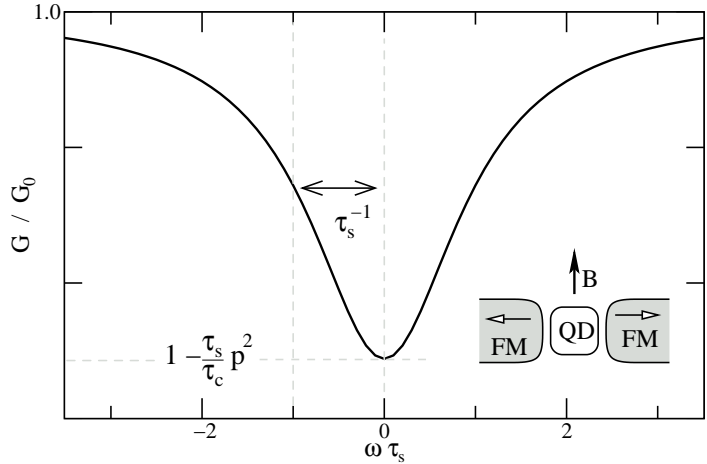}
\caption{\label{fig:G_lowp}
  Differential conductance, for ferromagnetic leads with anti-parallel
  magnetization, as a function of the magnetic field $\omega$ applied
  perpendicular to the accumulated spin. The half line width of the
  Hanle resonance directly determines the spin-decoherence time $\tau_s$.}
\end{center}
\end{figure}

We first focus on two ferromagnetic
leads with magnetization directions anti-parallel to each other,
see Fig.~\ref{fig:G_lowp}, and an arbitrary aligned external field.
The configuration has the advantage that the exchange field contributions
from the two leads cancel, and the spin dynamics is only govern by
the external field $\vec{B}_{\rm ext}$.
The linear conductance, then, is
\begin{eqnarray}
  \frac{G}{G_0} &=& 1-p^2\frac{\tau_{\rm s}}{\tau_{\rm c}} \,
  \frac{1+(\frac{\vec{m}_{\rm L}-\vec{m}_{\rm R}}{2}\vec{B}_{\rm ext}\tau_{\rm s})^2 } {1+(\vec{B}_{\rm ext} \tau_{\rm s})^2} \, .
\end{eqnarray}
where $G_0={e^2}\,{P_1}/{\tau_{\rm c}}{k_{\rm B}T}$ is
the asymptotic value of the conductance for a large magnetic field,
$|\vec{B}_{\rm ext}| \rightarrow \infty$, for which the spin accumulation is
completely destroyed. The latter is proportional to the single
occupation probability $P_1$.

If we assume the field to be aligned perpendicular to the lead
magnetizations (see Fig.~\ref{fig:G_lowp}), we find the Lorentzian
dependence on the external magnetic field that is familiar from the
optical Hanle effect.
The depth of the dip is given by $p^2\tau_{\rm s}/\tau_{\rm c}$ while
the width of the dip in Fig.~\ref{fig:G_lowp} provides a direct
access to the spin lifetime $\tau_{\rm s}$.
Of course, the conversion of applied magnetic field to frequency
requires the knowledge of the Lande factor $g$, which must be
determined separately like in Ref.~\cite{ralph}.

\subsubsection{Magnetic Field Applied Along $\vec{m}_{\rm L}+\vec{m}_{\rm R}$}
\label{subsubsec:noncollinear}

\begin{figure}[h!]
\begin{center}
\includegraphics[angle=-90,width=1.0\columnwidth]{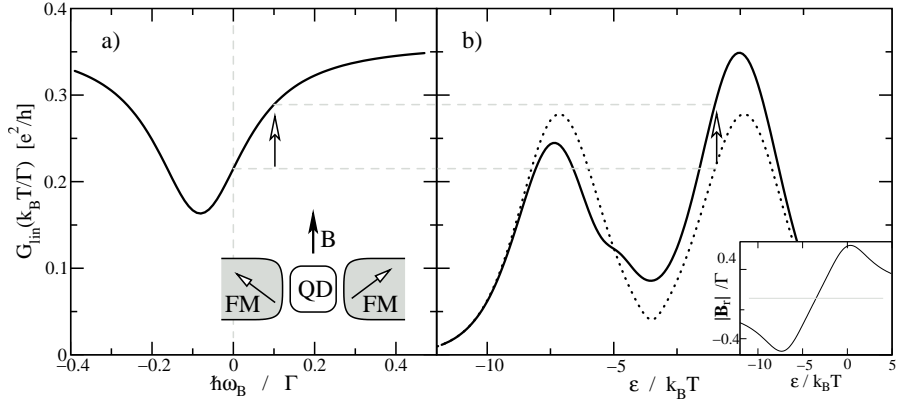}
\caption{\label{fig:G_eps}
  Linear conductance of the dot for an applied external magnetic
  field $\vec{B}_{\rm ext}$ along $\vec{m}_{\rm L}+\vec{m}_{\rm R}$.
  a) Linear conductance as a function of the applied field for
  $\varepsilon=0$.
  b) Linear conductance as a function of the level position
  $\varepsilon$ without external field (dotted) and for
  the applied field $|\vec{B}_{\rm ext}|=0.1\Gamma/\hbar$ (solid). 
  Further parameters are $\phi=3\pi/4$, $p=0.8$, $U=7k_{\rm B}T$, and
  $\tau_{\rm rel}=0$. The vertical lines relate the conductance
  increase of the dot at $\varepsilon=0$ for a magnetic field
  $\hbar\omega_{\rm B}=0.1\Gamma$.}
\end{center}
\end{figure}

Finally, we discuss the case of a non-collinear configuration
of the leads' magnetizations with a magnetic field applied along the direction
$\vec{m}_{\rm L}+\vec{m}_{\rm R}$ as shown in Fig.~\ref{fig:G_eps}.

In this case, both the exchange field and the external magnetic field are
pointing along the same direction $\vec{m}_{\rm L}+\vec{m}_{\rm R}$, so
their magnitude is just added. The linear conductance is, then,
\begin{eqnarray}
  \label{GB}
 \frac{G}{ G_0}  &=&
  1-p^2\frac{\tau_{\rm s}}{\tau_{\rm c}} \,\,
  \frac{\sin^2\frac{\phi}{2}}{1+(\vec{B}_{\rm ext}+\vec{B}_{\rm L}+\vec{B}_{\rm R})^2 \tau_{\rm s}^2}\, ,\qquad
\end{eqnarray}
where $\phi$ is the angle enclosed by $\vec{m}_{\rm L}$ and $\vec{m}_{\rm R}$.
The conductance as function of applied magnetic field as plotted in
Fig.~\ref{fig:G_eps}(a) reaches its minimal value when the sum of external
and exchange field vanishes. The exchange field leads to a
shift of the minimum's position  relative to $|\vec{B}_{\rm ext}|=0$ \cite{PRB}.
In real experiments, depending on the particular sample geometry,
one can expect a magnetic stray field, which is not considered to be part of
the experimentally applied magnetic field $\vec{B}_{\rm ext}$. These stray
fields also lead to a shift of the conductance minimum.
However, the analyzed setup of external field and magnetizations' directions 
allows for a stringent experimental 
verification of spin precession due to the exchange field.
To separate the exchange field from the influence of possible stray
fields its gate voltage dependence can be used.
The exchange interaction as function of the dot gate
voltage is plotted in the inset of Fig.~\ref{fig:G_eps}(b).
While the stray fields does not depend on gate voltage,
the exchange field does. In the flat band limit it even
changes sign as a function of gate voltage.
By plotting the conductance as function
of the gate voltage in Fig.~\ref{fig:G_eps}(b), we can observe the typical Coulomb
blockade oscillations, when the energy level of the empty or singly-occupied
dot becomes resonant with the lead Fermi energy. The interplay of exchange
and external field leads to an increase of conductance for one resonance
peak, but to a decrease for the other resonance.

\section{Conclusions}
\label{sec:conclusions}

We discussed the possibility to generate, manipulate, and probe single spins
in single-level quantum dots coupled to ferromagnetic leads.
A finite spin-polarization of the quantum-dot electron is achieved by
spin-polarized charge currents from or to the leads at finite bias voltage.
Any manipulation of the accumulated spin, e.g. by an external magnetic field
or by an intrinsic exchange field, is detectable in the electric current
through the device.
The occurrence of the exchange field is a consequence of many-body correlations
that are one of the intriguing features of nanostructures with large Coulomb
interaction.

We determine the dynamics of the quantum-dot spin by deriving expressions for
the spin currents through the tunnel barriers.
In addition to a contribution that is associated with the spin-polarization of
the charge currents from or to ferromagnets, there is a second contribution
describing transfer of angular momentum perpendicular to the leads' and dot's
magnetization that can be expressed in terms of the exchange field.

In order to manipulate the quantum-dot spin we suggest to make use of the
gate- and bias voltage dependence of the exchange field or to apply an
external magnetic field.
In particular, the spin precession modifies the dependence of the
linear conductance on the opening angle of the lead magnetizations.
The degree of modification is tunable by the gate voltage.
In nonlinear response, the bias-voltage dependence of the exchange field
can give rise to a negative differential conductance.
An application of a tunable external magnetic field allows one to determine the
dot-spin lifetime and to verify the existence of the intrinsic
spin precession caused by the exchange coupling.

\section*{Acknowledgments}
We thank J. Barna{\'s}, G. Bauer, A. Brataas, P. Brower, D. Davidovic, B. Kubala,
S. Maekawa, D. Ralph, G. Sch\"on, D. Urban, and B. Wunsch for
discussions.
This work was supported by the DFG under CFN, SFB 491, and GRK 726, the
EC RTN on 'Spintronics', Project PBZ/KBN/044/P03/2001 and the EC Contract
G5MACT-2002-04049.


\begin{thebibliography}{99.}

\bibitem{chargereview1}
D.V.~Averin, K.K.~Likharev: in {\it Mesoscopic Phenomenon in Solids},
ed. by B.L.~Altshuler, P.A.~Lee, R.A.~Webb (Amsterdam: North-Holland 1991)

\bibitem{chargereview2}
{\it Single Charge Tunneling: Coulomb Blockade Phenomena in
Nanostructures}, NATO ASI Series B: Physics 294, ed. by
H.~Grabert, M.H.~Devoret (Plenum Press, New York 1992)

\bibitem{chargereview3}
{\it Mesoscopic Electron Transport}, ed. by L.L.~Sohn,
L.P.~Kouwenhoven, G.~Sch\"on (Kluwer, Dordrecht 1997)


\bibitem{spinreview1}
S. A. Wolf, D. D. Awschalom, R. A. Buhrman, J. M. Daughton,
S. von Moln\'ar, M. L. Roukes, A. Y. Chtchelkanova, D. M. Treger,
Science, {\bf 294}, 1488-1495 (2001)

\bibitem{spinreview2}
I. Zutic, J. Fabian, and S. Das Sarma, Rev. Mod. Phys. {\bf 76}, 323 (2004)

\bibitem{Fulton}
T.~A. Fulton and G.~J. Dolan, Phys. Rev. Lett. {\bf 59}, 109 (1987)


\bibitem{GMR}
M.~N. Baibich, J. M. Broto, A. Fert, F. Nguyen Van Dau, F. Petroff,
P. Etienne, G. Creuzet, A. Friederich, and J. Chazelas,
Phys. Rev. Lett. {\bf 61}, 2472 (1988).


\bibitem{Julliere}
M. Julli{\`e}re, Phys. Lett. A {\bf 54}, 225 (1975).


\bibitem{Slonczewski}
J.~C. Slonczewski, Phys. Rev. B {\bf 39}, 6995 (1989).


\bibitem{Angular}
J.~S. Moodera and L.~R. Kinder,
J. Appl. Phys. {\bf 79}, 4724 (1996);
H. Jaffr{\`e}s, D. Lacour, F. Nguyen Van Dau, J. Briatico, F. Petroff, and A. Vaur\`es,
Phys. Rev. B {\bf 64}, 064427 (2001).




\bibitem{EPL}
M. Braun, J. K\"onig, and J. Martinek, Europhys. Lett. {\bf 72}, 294 (2005).


\bibitem{PRL}
J. K\"onig and J. Martinek, Phys. Rev. Lett. {\bf 90},  166602 (2003).

\bibitem{PRB}
M. Braun, J. K\"onig, and J. Martinek, Phys. Rev. B {\bf 70}, 195345 (2004).

\bibitem{Braig}
S. Braig and P. W. Brouwer, Phys. Rev. B 71, 195324 (2005).


\bibitem{Usaj}
G. Usaj and H. U. Baranger, Phys. Rev. B 63, 184418 (2001); G. Usaj
and H. U. Baranger, Phys. Rev. B 71, 179903 (E) (2005).




\bibitem{ono}
K. Ono, H. Shimada, and Y. Ootuka,
J. Phys. Soc. Jpn. {\bf 66}, 1261 (1997).



\bibitem{wees}
M. Zaffalon and B. J. van Wees,
Phys. Rev. Lett. {\bf91}, 186601 (2003).



\bibitem{granular}
L. F. Schelp, A. Fert, F. Fettar, P. Holody, S. F. Lee, J. L. Maurice,
F. Petroff, and A. Vaur\'es, Phys. Rev. B 56, R5747 (1997);
K. Yakushiji, S. Mitani, K. Takanashi, S. Takahashi, S. Maekawa,
H. Imamura, and H. Fujimori Appl. Phys. Lett.{\bf 78}, 515 (2001).

\bibitem{cnt}
A. Jensen, J. Nyg{\aa}rd and J. Borggreen in
{\it Proceedings of the International Symposium on Mesoscopic Superconductivity and Spintronics},
edited by H. Takayanagi and J. Nitta, (World Scientific 2003), pp. 33-37;
B. Zhao, I. M\"onch, H. Vinzelberg, T. M\"uhl, and C. M. Schneider,
Appl. Phys. Lett. {\bf 80}, 3144 (2002);
K. Tsukagoshi, B. W. Alphenaar, and H. Ago,
Nature {\bf 401}, 572 (1999).


\bibitem{ralph2}
A. N. Pasupathy, R. C. Bialczak, J. Martinek, J. E. Grose, L. A. K. Donev, P. L. McEuen, and D. C. Ralph,  Science, {\bf 306}, 86 (2004).


\bibitem{QD-exp}
Y. Chye, M. E. White, E. Johnston-Halperin, B. D. Gerardot, D. D. Awschalom, and P. M. Petroff,
Phys. Rev. B {\bf 66}, 201301(R) (2002).

\bibitem{ralph}
M. M. Deshmukh and D. C. Ralph,
Phys. Rev. Lett. {\bf 89}, 266803 (2002).


\bibitem{FMSTM}
 A. Kubetzka, M. Bode, O. Pietzsch, and R. Wiesendanger, Phys.
Rev. Lett. {\bf 88}, 057201 (2002).

\bibitem{schoenenberger}
S. Sahoo, C. Sch\"onenberger {\rm et al.}, Nature Physics {\bf 1}, 102 (2005).


\bibitem{grain_experiment}
L. Y. Zhang, C. Y. Wang, Y. G. Wei, X. Y. Liu, and D. Davidovi{\'c}, 
Phys. Rev. B {\bf 72}, 155445 (2005).


\bibitem{weymann}
I. Weymann, J. Barnas, J. K\"onig, J. Martinek, and G. Sch\"on,
Phys. Rev. B {\bf 72}, 113301 (2005);
I. Weymann, J. K\"onig, J. Martinek, J. Barnas, and G. Sch\"on,
Phys. Rev. B {\bf 72}, 115334 (2005).


\bibitem{kondo}
J. Martinek, Y. Utsumi, H. Imamura, J. Barna{\'s}, S. Maekawa, J. K\"onig, and G. Sch\"on,
Phys. Rev. Lett. {\bf 91}, 127203 (2003);
J. Martinek, M. Sindel, L. Borda, J. Barna{\'s}, J. K\"onig, G. Sch\"on, and J. von Delft,
Phys. Rev. Lett. {\bf 91}, 247202 (2003).



\bibitem{MS}
M. Braun, J. K\"onig, and J. Martinek, Superlat. and Microstruc. {\bf 37}, 333 (2005).


\bibitem{MeirWingreen}
Y. Meir and N. S. Wingreen, Phys. Rev. Lett. {\bf 68}, 2512 (1992).


\bibitem{spinmix}
A. Brataas, Y.~V. Nazarov, and G.~E.~W. Bauer,
Eur. Phys. J B {\bf 22}, 99 (2001).





\bibitem{bauer}
A. Brataas, Y. Tserkovnyak, G.~E.~W. Bauer, and B.~I. Halperin, Phys. Rev. B {\bf66}, 060404(R) (2002).


\bibitem{Springer}
J. K\"onig, J. Martinek, J. Barnas, and G. Sch\"on,
in "CFN Lectures on Functional Nanostructures", Eds. K. Busch {\em et al.},
Lecture Notes in Physics {\bf 658}, Springer, 145-164 (2005).

\bibitem{spinblock1}
D. Weinmann, W. H\"ausler, and B. Kramer,
Phys. Rev. Lett. {\bf 74}, 984 (1995);
A.~K. Huettel, H. Qin, A.~W. Holleitner, R.~H. Blick,
K. Neumaier, D. Weinmann, K. Eberl, and J.~P. Kotthaus,
Europhys. Lett. {\bf 62}, 712 (2003).

\bibitem{spinblock2}
K. Ono, D. G. Austing, Y. Tokura, and S. Tarucha,
Science {\bf 297}, 1313 (2002).


\bibitem{johnson}
M. Johnson and R.~H. Silsbee,
Phys. Rev. Lett. {\bf 55}, 1790 (1985); Phys. Rev. B. {\bf 37}, 5326 (1988).

\bibitem{hanle_in_dots}
R.~J. Epstein,
D.~T. Fuchs, W.~V. Schoenfeld, P.~M. Petroff, and D.~D. Awschalom,
Appl. Phys. Lett. {\bf 78}, 733 (2001).


\end{thebibliography}

%



\printindex


\end{document}